\documentclass[sigconf,table]{acmart}
\usepackage{makecell}
\usepackage{booktabs}
\usepackage{multirow}
\usepackage{subcaption}
\usepackage{braket}
\usepackage{minted}
\usepackage{float}
\usepackage{multicol}

\setcopyright{none}
\renewcommand\footnotetextcopyrightpermission[1]{} 

\acmConference[SBLP'25]{XXIX Brazilian Symposium on Programming Languages}{September 22--26, 2025}{Recife, PE}

\begin{document}

\title{Quantum Gate Decomposition}
\subtitle{A Study of Compilation Time vs. Execution Time Trade-offs}

\author{Evandro C. R. Rosa}
\orcid{0000-0002-8197-9454}

\affiliation{%
    \institution{Departamento de Informática e Estatística,\\Universidade Federal de Santa Catarina}
    \city{Florianópolis}
    \country{Brazil}
}
\email{evandro.crr@posgrad.ufsc.br}

\author{Jerusa Marchi}
\orcid{0000-0002-4864-3764}

\affiliation{%
    \institution{Departamento de Informática e Estatística,\\Universidade Federal de Santa Catarina}
    \city{Florianópolis}
    \country{Brazil}
}
\email{jerusa.marchi@ufsc.br}

\author{Eduardo I. Duzzioni}
\orcid{0000-0002-8971-2033}

\affiliation{%
    \institution{Departamento de Física,\\Universidade Federal de Santa Catarina}
    \city{Florianópolis}
    \country{Brazil}
}
\email{eduardo.duzzioni@ufsc.br}

\author{Rafael de Santiago}
\orcid{0000-0001-7033-125X}

\affiliation{%
    \institution{Departamento de Informática e Estatística,\\Universidade Federal de Santa Catarina}
    \city{Florianópolis}
    \country{Brazil}
}
\email{r.santiago@ufsc.br}

\begin{abstract}
    Similar to classical programming, high-level quantum programming languages generate code that cannot be executed directly by quantum hardware and must be compiled. However, unlike classical code, quantum programs must be compiled before each execution, making the trade-off between compilation time and execution time particularly significant.
    In this paper, we address the first step of quantum compilation: multi-qubit gate decomposition. We analyze the trade-offs of state-of-the-art decomposition algorithms by implementing them in the Ket quantum programming platform and collecting numerical performance data. This is the first study to both implement and analyze the current state-of-the-art decomposition methods within a single platform.
    Based on our findings, we propose two compilation profiles: one optimized for minimizing compilation time and another for minimizing quantum execution time. Our results provide valuable insights for both quantum compiler developers and quantum programmers, helping them make informed decisions about gate decomposition strategies and their impact on overall performance.
\end{abstract}

\keywords{Quantum Computing, Quantum Programming, Quantum Compiler, Gate Decomposition, Ket}

\renewcommand\abstractname{ABSTRACT}
\renewcommand\keywordsname{KEYWORDS}

\renewcommand\refname{REFERENCES}

\maketitle

\section{Introduction}

For programming languages and platforms that treat the quantum bit (qubit) as a first-class citizen in quantum programming, \textit{e.g.}, Q\#~\cite{svoreEnablingScalableQuantum2018} and Ket~\cite{darosaKetQuantumProgramming2022}, coding a quantum application involves invoking functions known as quantum gates. Quantum gates have no side effects on the classical state of the program; they only affect the quantum state and can induce superposition and entanglement~\cite{nielsenQuantumComputationQuantum2010}. Another class of functions, known as measurements, is used to extract information from the quantum state, returning it as classical data while causing the collapse of the quantum state, \emph{i.e.}, destroying the superposition.



\begin{figure}[htbp]
    \centering
    \begin{subfigure}[b]{.49\linewidth}
        \begin{minted}[frame=lines,fontsize=\footnotesize,breaklines]{py}
def diffusion(q : Quant):
    with around(cat(H, X), q):
        with control(q[:-1]):
            Z(q[-1])
        \end{minted}
        \caption{Ket code.}
        \label{fig:ket:diffusion}
    \end{subfigure}
    \hfil
    \begin{subfigure}[b]{.35\linewidth}
        \centering
        \includegraphics[width=.78\linewidth]{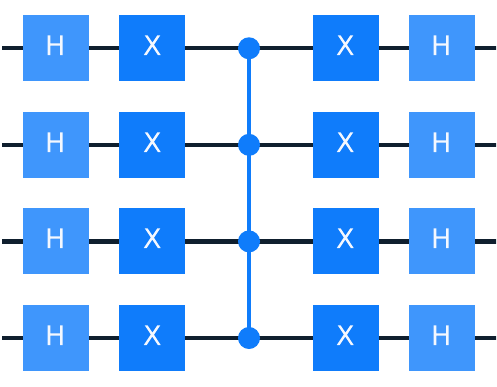}
        \caption{Quantum Circuit.}
        \label{fig:cirq:diffusion}
    \end{subfigure}

    \begin{subfigure}[b]{\linewidth}
        \includegraphics[width=\linewidth]{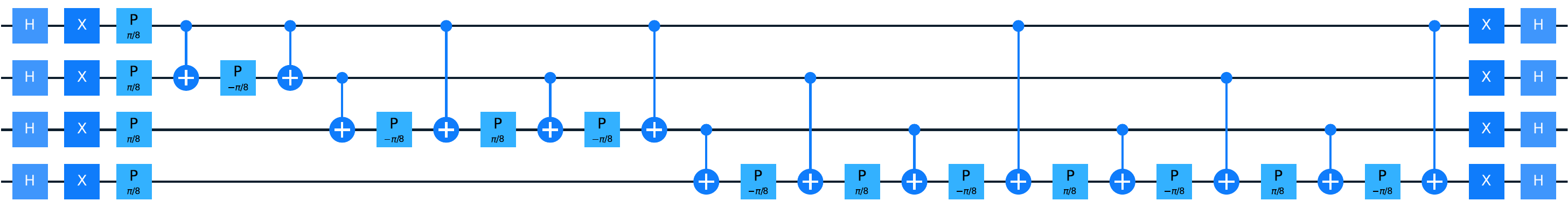}

        \caption{Decomposed Circuit.}
        \label{fig:cirq:diffusion_decompose}
    \end{subfigure}
    \caption{Grover diffusion operation implemented using Ket~(\ref{sub@fig:ket:diffusion}), along with its corresponding high-level~(\ref{sub@fig:cirq:diffusion}) and decomposed~(\ref{sub@fig:cirq:diffusion_decompose}) quantum circuit for 4 qubits.}
    \Description{}
\end{figure}

Figure~\ref{fig:ket:diffusion} illustrates a simple quantum program written in Python using the Ket quantum programming platform~\cite{darosaKetQuantumProgramming2022}. This code implements the Grover diffusion operator, a key component of Grover's quantum search algorithm~\cite{groverFastQuantumMechanical1996}, utilizing the Hadamard~(\texttt{H}), Pauli~\texttt{X}, and Pauli~\texttt{Z} gates. The term \emph{quantum gate}, or more precisely \emph{quantum logical gate}, is inspired by classical circuit and logic gates. Each quantum code has an equivalent quantum circuit; for example, the circuit in Figure~\ref{fig:cirq:diffusion} is equivalent to the code in Figure~\ref{fig:ket:diffusion} applied for 4-qubits.

Despite the direct use of quantum bits and quantum gates in quantum programming, which gives the impression of low-level programming, the approach provides feature-rich high-level programming~\cite{rosaOptimizingGateDecomposition2025}, producing operations that cannot be executed directly by a quantum computer and must undergo a compilation process. For example, in Figure~\ref{fig:cirq:diffusion}, note that the quantum gate in the middle of the circuit, a multi-controlled Pauli~\texttt{Z} gate, encompasses all 4 qubits. This type of multi-qubit gate cannot be executed directly by a quantum computer and therefore must be decomposed into a sequence of one- and two-qubit gates with equivalent effect, as shown in Figure~\ref{fig:cirq:diffusion_decompose}. This is analogous to a line of high-level classical code being decomposed into several lines of assembly code.

The quantum compilation process can be split into three main parts: quantum gate decomposition~\cite{dasilvaLineardepthQuantumCircuits2022,itenIntroductionUniversalQCompiler2021,rosaOptimizingGateDecomposition2025,valeCircuitDecompositionMulticontrolled2024,barencoElementaryGatesQuantum1995,itenQuantumCircuitsIsometries2016,rosaAutomatedAuxiliaryQubit2024,zindorfEfficientImplementationMultiControlled2024}, where multi-qubit gates are broken down into a sequence of one- and two-qubit operations; circuit mapping~\cite{chowdhuryQubitAllocationStrategies2024,liTacklingQubitMapping2019,niuHardwareAwareHeuristicQubit2020,willeMQTQMAPEfficient2023,zhuDynamicLookAheadHeuristic2020,zhuVariationAwareQuantumCircuit2023}, where the logical qubits are mapped to physical qubits that are limited in their connectivity, thereby restricting which qubits can participate in a two-qubit gate; and lastly, a pulse schedule is generated based on the mapped circuit~\cite{huangCalibratingSinglequbitGates2023,stefanazziQICKQuantumInstrumentation2022,alexanderQiskitPulseProgramming2020}. These pulses are then passed to an arbitrary waveform generator (AWG) that physically controls the qubits. Each step of this decomposition process brings the quantum code closer to the hardware execution and becomes more hardware-dependent.

This paper addresses the first step of the quantum compiler, namely the quantum gate decomposition, by providing an analysis of the trade-offs between different quantum gate decomposition algorithms in terms of compilation time and execution time. Unlike classical compilation, where code is compiled once and executed many times, quantum compilation often occurs during runtime, making this trade-off particularly significant in quantum programming.

We implemented state-of-the-art quantum gate decomposition algorithms in Ket's quantum compiler, Libket, and evaluated the compilation and execution times of each algorithm. There is no definitive answer as to whether prioritizing execution time over compilation time, or \emph{vice-versa}, is preferable; the best choice depends on the specifications of the classical and quantum hardware. The objective of this paper is to provide a basis for helping quantum compiler developers and programmers make informed decisions.

The main contributions of this paper are:
\begin{itemize}
    \item A comprehensive survey of quantum gate decomposition algorithms, including optimizations beyond those proposed by the original authors.
    \item  Implementation of all surveyed algorithms in the Ket quantum compiler--marking the first unified implementation on a single platform and enabling consistent benchmarking.
    \item A classification of the algorithms into two categories, forming the basis for two quantum compilation profiles: one focused on minimizing compilation time, the other on reducing quantum execution time.
\end{itemize}

The remainder of this paper is organized as follows.
Section~\ref{sec:runtime} introduces classical and quantum runtime, highlighting the implications for quantum program compilation. Section~\ref{sec:hlqp} examines high-level quantum programming and shows how multi-qubit gates naturally arise in such code. Section~\ref{sec:algorithms} then surveys the state-of-the-art decomposition algorithms used to break down these gates into executable instructions. Section~\ref{sec:benchmark} follows with a performance evaluation of these algorithms, focusing on CNOT count and circuit depth. Based on the benchmark data, Section~\ref{sec:results} analyzes the trade-offs involved and introduces two quantum compilation profiles. Finally, Section~\ref{sec:conclusion} presents our conclusions and final remarks.

\section{Classical and Quantum Runtime}
\label{sec:runtime}

A quantum program is not entirely quantum; rather, it is a classical-quantum program. Classical processing is always required at least to prepare the quantum circuit inputs and process the quantum execution outputs. The quantum computer can be seen as an accelerated processing unit, similar to an FPGA or GPU, where the CPU manages the execution. This implies that a quantum application has two distinct runtimes: a classical runtime, which includes all program execution, and a quantum runtime, which occurs within it, with a program potentially initiating several quantum runtimes.

Unlike classical code, which can be compiled once and reused with different inputs, a quantum circuit must be compiled separately for each specific input. This means that a quantum program designed to work with varying inputs can only be constructed at classical runtime--after all necessary parameters are known. As a result, the quantum portion of a classical-quantum program cannot be compiled in advance.

For instance, in Figure~\ref{fig:ket:diffusion}, the function \texttt{diffusion} takes a list of qubits as input. In classical compilation, this would typically generate a loop whose size depends on the input list. However, for quantum compilation, the exact number of qubits must be known in advance to generate the correct quantum circuit. This requirement means the quantum compiler must be invoked dynamically at classical runtime--just before executing the quantum code.

The performance of a quantum application is evaluated by the number of two-qubit gates in the compiled quantum circuit. This two-qubit gate is typically a CNOT gate\footnote{All analyses in this paper are also valid if CZ gates are used instead.}. Single-qubit gates are generally not counted, as sequences of such gates can often be merged into a single equivalent operation. The time required to execute quantum code on a quantum computer is directly related to the circuit depth. Since gates acting on independent qubits can be executed in parallel, the execution time may be shorter than the total number of CNOT gates. For this reason, circuit depth serves as a metric for quantum execution time. Conversely, the number of CNOTs impacts compilation time, as each CNOT represents a computational step for the quantum compiler.

Since the compilation of the quantum circuit occurs at classical runtime, the trade-off between compilation time and quantum execution time directly impacts the overall classical-quantum execution time. This trade-off can also be viewed as a balance between classical computation (compilation) and quantum execution time.

\section{High-Level Quantum Programming}
\label{sec:hlqp}

While quantum programming introduces unique challenges--such as the impossibility of copying quantum data due to the no-cloning theorem~\cite{woottersSingleQuantumCannot1982}--it also offers powerful abstractions that simplify the development of quantum applications. One such feature is the ability to automatically invoke the inverse of a quantum operation. In this paper, we examine quantum programming through the lens of Ket\footnote{\url{https://quantumket.org}}, an open-source quantum programming platform with a Python API~\cite{darosaKetQuantumProgramming2022}.

At its core, Ket provides a minimal yet expressive set of quantum gates: the Pauli gates~(\texttt{X},~\texttt{Y},~and~\texttt{Z}); Rotation gates~(\texttt{RX},~\texttt{RY},~and~\texttt{RZ}); the Phase gate~(\texttt{P}); and the Hadamard gate~(\texttt{H}). These gates are sufficient to prepare a qubit in any desired state. However, on their own, they cannot achieve universal quantum computation, as single-qubit gates cannot generate entanglement. To address this limitation, Ket allows any gate to be applied with one or more control qubits, enabling the creation of entangled states.

By adding just a controlled NOT\footnote{Also known as the controlled Pauli~\texttt{X} or simply the CNOT gate.} to the basic set of quantum gates, universal quantum computation becomes achievable. In Ket, the CNOT gate is provided by a function of the same name, but it can also be constructed using the function \texttt{ctrl} as
\begin{equation*}
    \texttt{lambda~c,~t:~ctrl(c,~X)(t)},
\end{equation*}
or by using the ``\texttt{with~control}'' construction. Any quantum gate or function that contains quantum gates--as in the one shown in Figure~\ref{fig:ket:diffusion}--can be called with control qubits. This flexibility to append control qubits to any quantum gate greatly enhances high-level quantum programming by enabling the easy construction of complex operations from basic quantum gates.

A controlled quantum gate call is the quantum analog of the classical \texttt{if} statement, where a gate is applied to the target qubits only if the control qubits are all in the state $\ket{1}$. The key difference is that a controlled operation acts on the superposition and has the ability to create entanglement.

The code in Figure~\ref{fig:ket:state} illustrates how controlled operations can appear in high-level quantum programming. The function \texttt{prepare} takes a list of qubits alongside a list of measurement probabilities $[r_k]$ and a list of phase values $[\theta_k]$, and prepares the qubits in the state $\sum_k \sqrt{r_k}\,e^{i\theta_k}\ket{k}$. This function calls itself recursively, adding a control qubit at each recursion. Therefore, even though there is no direct call for controlled \texttt{RY} or \texttt{P} gates, the execution of this code creates several controlled operations.

\begin{figure}[htbp]
    \centering
    \begin{minipage}{.72\linewidth}
        \begin{minted}[frame=lines,fontsize=\footnotesize,breaklines]{py}
def prepare(
    qubits: Quant,
    prob: ParamTree | list[float],
    amp: list[float] | None = None,
):
    if not isinstance(prob, ParamTree):
        prob = ParamTree(prob, amp)
    head, *tail = qubits
    RY(prob.value, head)
    if prob.is_leaf():
        with around(X, head):
            P(prob.phase0, head)
        return P(prob.phase1, head)
    with around(X, head):
        ctrl(head, prepare)(tail, prob.left)
    ctrl(head, prepare)(tail, prob.right)
        \end{minted}
    \end{minipage}
    \caption{Arbitrary quantum state preparation algorithm implemented using Ket. For the implementation of the \texttt{ParamTree}, see reference \citet[Figure 10a]{rosaOptimizingGateDecomposition2025}.}
    \label{fig:ket:state}
    \Description{}
\end{figure}

Alongside controlled gate calls, Ket provides constructions that take advantage of the reversibility of quantum computation. A key example is the ``\texttt{with~around}'' construction, which acts as a basis change--analogous to basis changes in linear algebra--for the inner scope. In the code of Figure~\ref{fig:ket:state}, the \texttt{with~around} construction is used to change the control state from $\ket{1}$ to $\ket{0}$. This construction also allows the compiler to remove some controlled gate calls, for the code from Figure~\ref{fig:ket:state}, those associated with the \texttt{X} gate~\cite{rosaOptimizingGateDecomposition2025}.

\section{Decomposition Algorithms}
\label{sec:algorithms}

\begin{figure*}[htbp]
    \centering
    \begin{minipage}[b]{.65\linewidth}
        \begin{table}[H]
            \scriptsize
            \caption{Gate decomposition algorithms evaluated in this paper. \emph{No. Aux} denotes the number of auxiliary qubits required, and \emph{Aux State} specifies their state. \emph{Depth}, \emph{No. CNOTs}, and \emph{No. Aux} are reported for an $n$-controlled gate decomposition.}
            \label{tab:algorithms}
            \begin{tabular}{cllcccc}
                \toprule
                Gates                               & Algorithms                       & Mode                      & No. Aux                                      & Aux State                      & Depth                             & No. CNOTs                          \\
                \midrule
                \midrule
                \multirow{7}{*}{Pauli Gates}        & \multirow{4}{*}{V Chain}         & \multirow{2}{*}{C2X}      & \multirow{2}{*}{$n-2$}                       & \cellcolor{gray!10}Clean       & \cellcolor{gray!10}$4n$           & \cellcolor{gray!10}$6n$            \\
                                                    &                                  &                           &                                              & Dirty                          & $8n$                              & $8n$                               \\
                \cmidrule(lr){3-7}
                                                    &                                  & \multirow{2}{*}{C3X}      & \multirow{2}{*}{$\lceil\frac{n-3}{2}\rceil$} & \cellcolor{gray!10}Clean       & \cellcolor{gray!10}$6n$           & \cellcolor{gray!10}$6n$            \\
                                                    &                                  &                           &                                              & Dirty                          & $12n$                             & $12n$                              \\
                \cmidrule(lr){2-7}
                                                    & \multirow{3}{*}{Single Aux }     & \cellcolor{gray!10}Linear & \cellcolor{gray!10}$1$                       & \cellcolor{gray!10}Clean/Dirty & \cellcolor{gray!10}$8n$           & \cellcolor{gray!10}$12n$           \\
                                                    &                                  & \multirow{2}{*}{Log}      & \multirow{2}{*}{$1$}                         & Clean                          & $O(\log(n)^3)$                    & $O(n\log(n)^4)$                    \\
                                                    &                                  &                           &                                              & \cellcolor{gray!10}Dirty       & \cellcolor{gray!10}$O(\log(n)^3)$ & \cellcolor{gray!10}$O(n\log(n)^4)$ \\
                \cmidrule(lr){1-7}
                \multirow{2}{*}{Rotation Gates}     & \multirow{2}{*}{SU(2)}           & \cellcolor{gray!10}Linear & \cellcolor{gray!10}0                         & \cellcolor{gray!10}N/A         & \cellcolor{gray!10}$8n$           & \cellcolor{gray!10}$12n$           \\
                                                    &                                  & Log                       & 0                                            & N/A                            & $O(\log(n)^3)$                    & $O(n\log(n)^4)$                    \\
                \cmidrule(lr){1-7}
                \multirow{2}{*}{Phase and Hadamard} & \cellcolor{gray!10}SU(2) Rewrite & \cellcolor{gray!10}       & \cellcolor{gray!10}$1$                       & \cellcolor{gray!10}Clean       & \cellcolor{gray!10}$8n$           & \cellcolor{gray!10}$12n$           \\
                                                    & Single Aux U(2)                  &                           & $1$                                          & Clean                          & $O(\log(n)^3)$                    & $O(n\log(n)^4)$                    \\
                \cmidrule(lr){1-7}
                \multirow{3}{*}{U(2)}               & \multirow{2}{*}{Network}         & \cellcolor{gray!10}C2X    & \cellcolor{gray!10}$\approx n$               & \cellcolor{gray!10}Clean       & \cellcolor{gray!10}$6\log_2(n)$   & \cellcolor{gray!10}$6n$            \\
                                                    &                                  & C3X                       & $\approx \frac{n}{2} $                       & Clean                          & $8\log_2(n)$                      & $6n$                               \\
                                                    & \cellcolor{gray!10}Liner Depth   & \cellcolor{gray!10}       & \cellcolor{gray!10}0                         & \cellcolor{gray!10}N/A         & \cellcolor{gray!10}$16n$          & \cellcolor{gray!10}$n^2/10$        \\

                \bottomrule
            \end{tabular}
        \end{table}
    \end{minipage}
    \hfil
    \begin{minipage}[b]{.3\linewidth}
        \begin{figure}[H]
            \centering
            \begin{subfigure}[b]{.49\linewidth}
                \centering
                \includegraphics[height=154px]{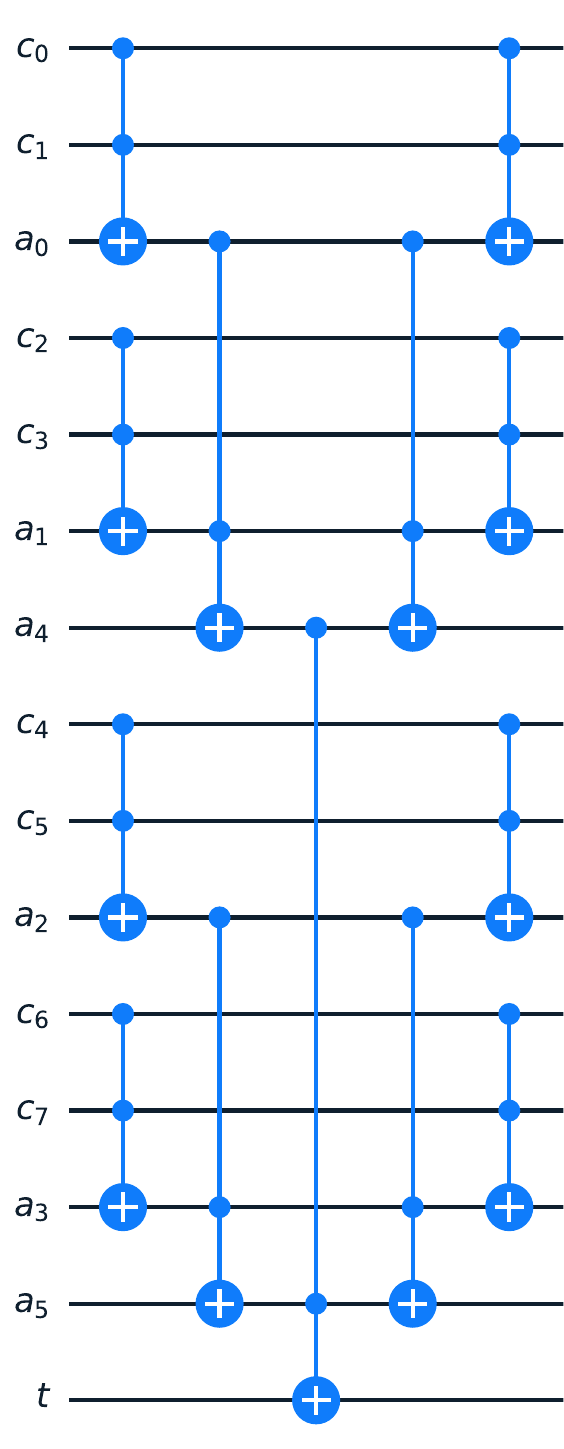}
                \caption{Network C2X.}
                \label{fig:network:c2x}
            \end{subfigure}
            \hfil
            \begin{subfigure}[b]{.49\linewidth}
                \centering
                \includegraphics[height=154px]{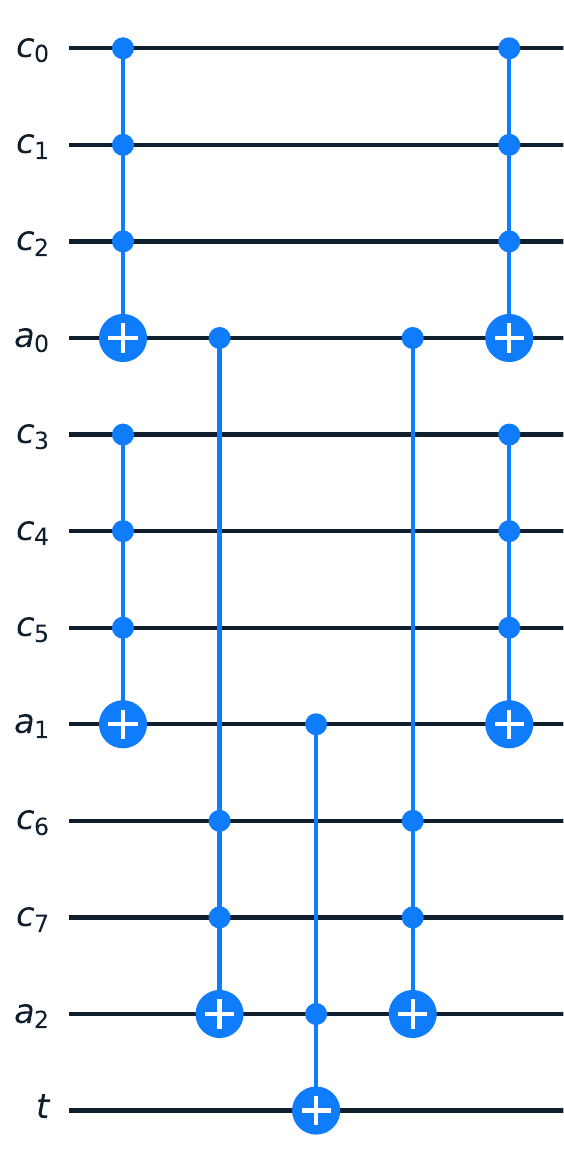}
                \caption{Network C3X.}
                \label{fig:network:c3x}
            \end{subfigure}
            \caption{Network decomposition for {8-controlled} Pauli~\texttt{X}.}
            \label{fig:network}
            \Description{}
        \end{figure}
    \end{minipage}
\end{figure*}

In Ket, multi-qubit gates are constructed such that only multi-controlled versions of the gates \texttt{X}, \texttt{Y}, \texttt{Z}, \texttt{RX}, \texttt{RY}, \texttt{RZ}, \texttt{P},~and~\texttt{H} arise from high-level programming. This constraint allows the quantum compiler to implement a streamlined and efficient set of decomposition algorithms. We categorize these gates into three distinct groups based on their decomposition requirements: (i) Pauli Gates (\texttt{X}, \texttt{Y}, and \texttt{Z}), a decomposition algorithm for one of these gates can be applied to the others with a simple basis change on the target qubit; (ii) Rotation Gates (\texttt{RX}, \texttt{RY}, and \texttt{RZ}), these belong to the special unitary group SU(2); (iii) Phase and Hadamard Gates (\texttt{P} and \texttt{H}), these belong to the broader unitary group U(2)\footnote{All single-qubit gates are in U(2), with Rotation gates in SU(2) $\subset$ U(2).}.

Table~\ref{tab:algorithms} summarizes state-of-the-art quantum gate decomposition algorithms. Some of these algorithms require \emph{auxiliary qubits}, which are additional qubits beyond those directly involved in the controlled gate. Depending on the algorithm, auxiliary qubits must either be initialized in the $\ket{0}$ state (Clean auxiliary) or can be in any state (Dirty auxiliary). Regardless of their initial state, auxiliary qubits must be restored to their original state after decomposition so that they can be reused elsewhere and do not create new entanglements. Typically, decompositions using more auxiliary and/or clean qubits are more efficient. The Ket compiler automatically manages the allocation and usage of auxiliary qubits when available~\cite{rosaAutomatedAuxiliaryQubit2024}.

Table~\ref{tab:algorithms} also presents algorithm variations, such as C2X and C3X, which use approximations of 2-controlled and 3-controlled NOT gates~\cite{maslovAdvantagesUsingRelativephase2016}, respectively. Additionally, some algorithms are labeled \emph{Linear} or \emph{Log}, which are algorithms that share similar auxiliary requirements.

The quantum circuit depth and the number of CNOTs presented in Table~\ref{tab:algorithms} were obtained by fitting the curves of the benchmark data. In cases where a good curve fit was not possible, we present the complexity as stated by the authors in big-O notation.

In the following subsections, we discuss the decomposition algorithms required for each class of quantum gates. In Section~\ref{sec:benchmark}, we present a performance comparison of the decomposition algorithms.

\subsection{Pauli Gates}

\begin{figure}[htbp]
    \centering
    \begin{subfigure}[b]{.37\linewidth}
        \begin{minted}[frame=lines,fontsize=\footnotesize,breaklines]{py}
def myY(q):
    with around(SD, q):
        X(q)
def myZ(q):
    with around(H, q):
        X(q)

def myCY(c, t):
    ctrl(c, myY)(t)
def myCZ(c, t):
    ctrl(c, myZ)(t)
        \end{minted}
        \caption{Ket code.}
        \label{fig:ket:myGate}
    \end{subfigure}
    \hfil
    \begin{minipage}[b]{.5\linewidth}
        \centering
        \begin{subfigure}[c]{\linewidth}
            \centering
            \includegraphics[height=43px]{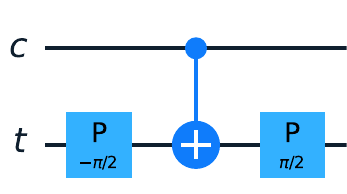}
            \caption{\texttt{myCY} Circuit.}
            \label{fig:cirq:myCY}
        \end{subfigure}

        \begin{subfigure}[b]{\linewidth}
            \centering
            \includegraphics[height=43px]{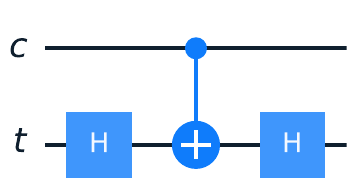}
            \caption{\texttt{myCZ} Circuit.}
            \label{fig:cirq:myCZ}
        \end{subfigure}
    \end{minipage}

    \caption{Defining the gates \texttt{Y} and \texttt{Z} using the \texttt{X} gate and a basis change, allowing the controlled \texttt{X} gate to implement the controlled versions of these gates.}
    \Description{}
\end{figure}

The Pauli gates are among the most commonly used quantum gates in algorithm design. A key insight for gate decomposition is that the decomposition of one Pauli gate can be adapted for the others. For example, in Figure~\ref{fig:ket:myGate}, the \texttt{Y} and \texttt{Z} gates are defined using the \texttt{X} gate with a basis change. Consequently, the circuits for their controlled versions, shown in Figures~\ref{fig:cirq:myCY} and~\ref{fig:cirq:myCZ}, rely on a controlled \texttt{X} gate. This example explicitly illustrates the behavior of Ket's compiler, which automatically applies this principle: only the decomposition of the Pauli \texttt{X} gate is explicitly defined, while the other Pauli gates are implemented through basis changes. In this paper, we evaluate only the performance of the \texttt{X} gate decomposition, as the decomposition of the other Pauli gates requires only an additional two single-qubit gates.

\subsubsection*{Network}

The Network decomposition~\cite[p. 183]{nielsenQuantumComputationQuantum2010} is one of the most efficient algorithms and can be applied to any multi-controlled gate. Figure~\ref{fig:network} illustrates two variants of the algorithm for a 8-controlled Pauli \texttt{X} gate. Note that, in addition to the target and control qubits, this decomposition requires auxiliary qubits, all initialized to the $\ket{0}$ state. For Pauli gate decomposition, we can reduces the number of auxiliary qubits by one or two compared to other gates.

This decomposition results in a circuit depth $O(\log(n))$ for an $n$-controlled gate when organizing the gates as presented by \citet[Corollary 5]{maslovAdvantagesUsingRelativephase2016}. Additionally, for the decomposition of 2- and 3-controlled \texttt{X} gates, where the original target qubit is not directly affected, approximate versions of this decomposition are applied~\cite{maslovAdvantagesUsingRelativephase2016}.

\subsubsection*{V Chain}

\begin{figure}[htbp]
    \centering
    \begin{subfigure}[b]{.6\linewidth}
        \centering
        \includegraphics[height=94px]{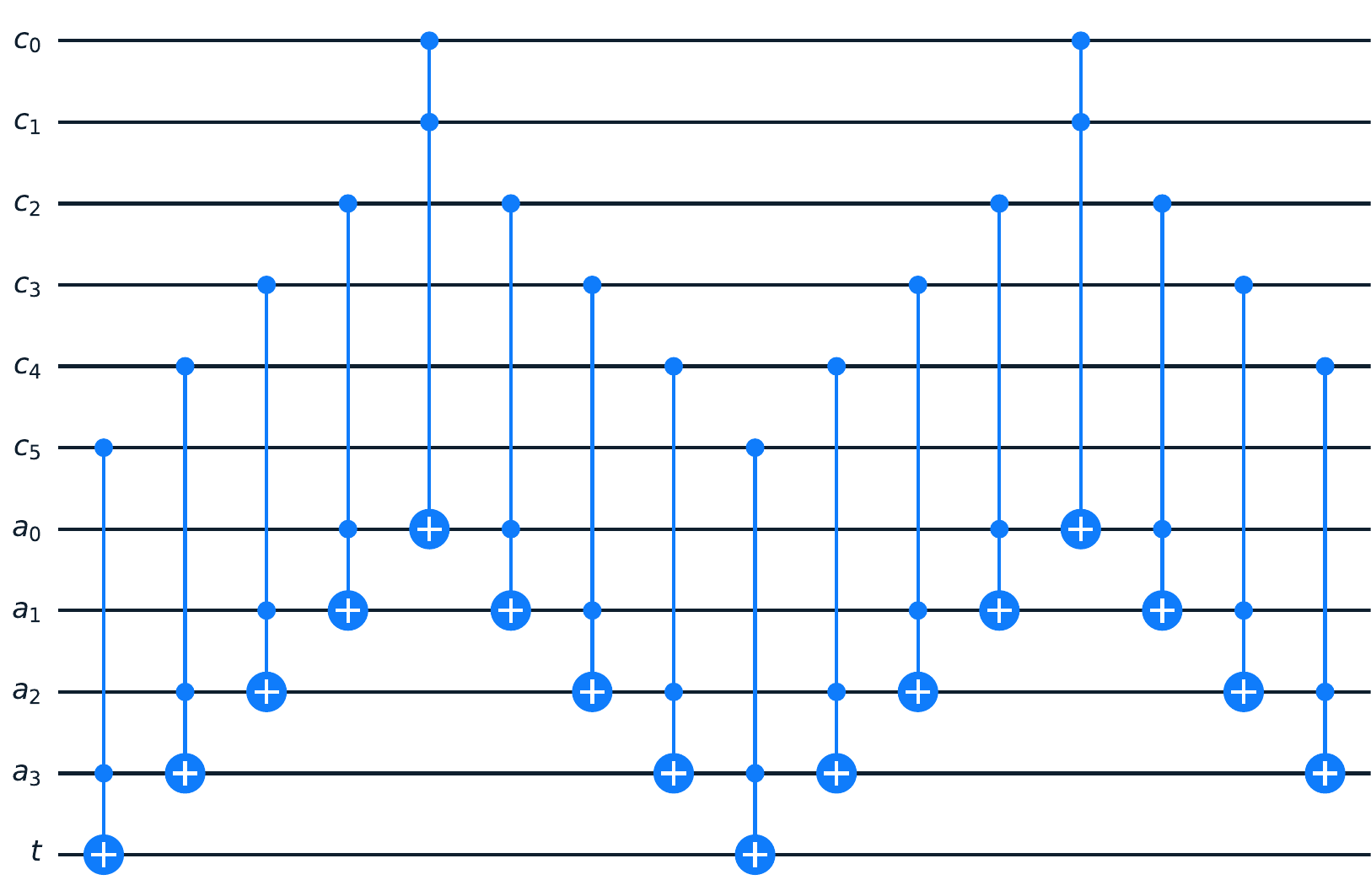}
        \caption{V Chain C2X, Dirty.}
    \end{subfigure}
    \hfil
    \begin{subfigure}[b]{.39\linewidth}
        \centering
        \includegraphics[height=94px]{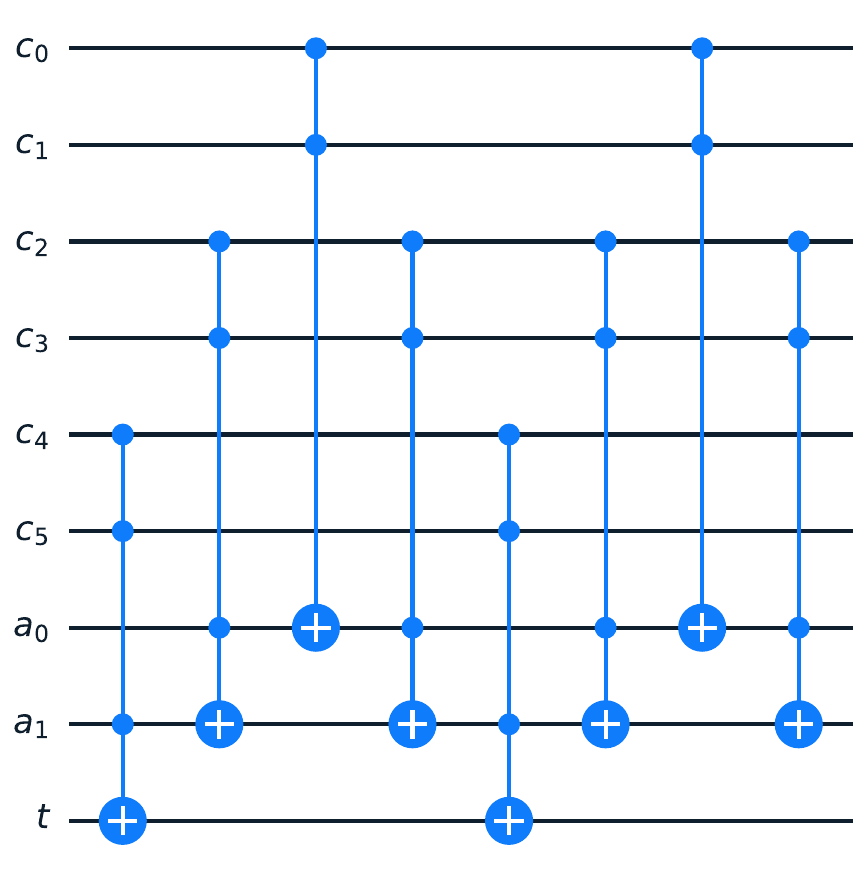}
        \caption{V Chain C3X, Dirty.}
    \end{subfigure}
    \caption{V Chain decomposition for 6-controlled Pauli~\texttt{X}.}
    \label{fig:vchain}
    \Description{}
\end{figure}

The V Chain decomposition~\cite{barencoElementaryGatesQuantum1995} has a structure similar to the Network decomposition for Pauli gates but allows the use of either dirty or clean auxiliary qubits. Like the Network decomposition, the V Chain decomposition utilizes approximate 2- or 3-controlled \texttt{X}. Although this algorithm results in a number of CNOT gates similar to the Network decomposition, especially when using clean auxiliary qubits, it produces a circuit with linear depth rather than logarithmic depth. Figure~\ref{fig:vchain} illustrates a V Chain decomposition using dirty auxiliary qubits for a 6-controlled \texttt{X} gate.

\subsubsection*{Single Aux}

The Single Aux decomposition algorithms require only a single auxiliary qubit and include two methods: one that results in a quantum circuit with linear depth, proposed by \citet{zindorfEfficientImplementationMultiControlled2024}, and another that achieves logarithmic depth, proposed by \citet{claudonPolylogarithmicdepthControlledNOTGates2024}.

\begin{figure}[htbp]
    \centering
    \begin{subfigure}[b]{.3\linewidth}
        \centering
        \includegraphics[height=110px]{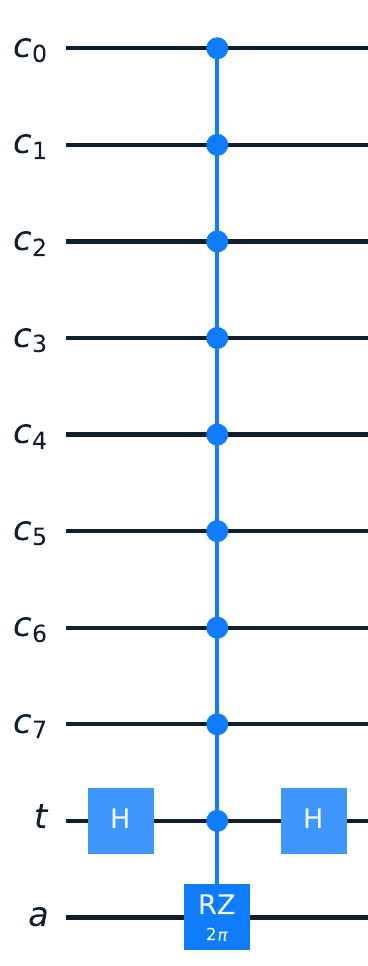}
        \caption{Linear, Dirty.}
        \label{fig:single_aux:linear}
    \end{subfigure}
    \hfil
    \begin{subfigure}[b]{.65\linewidth}
        \centering
        \includegraphics[height=110px]{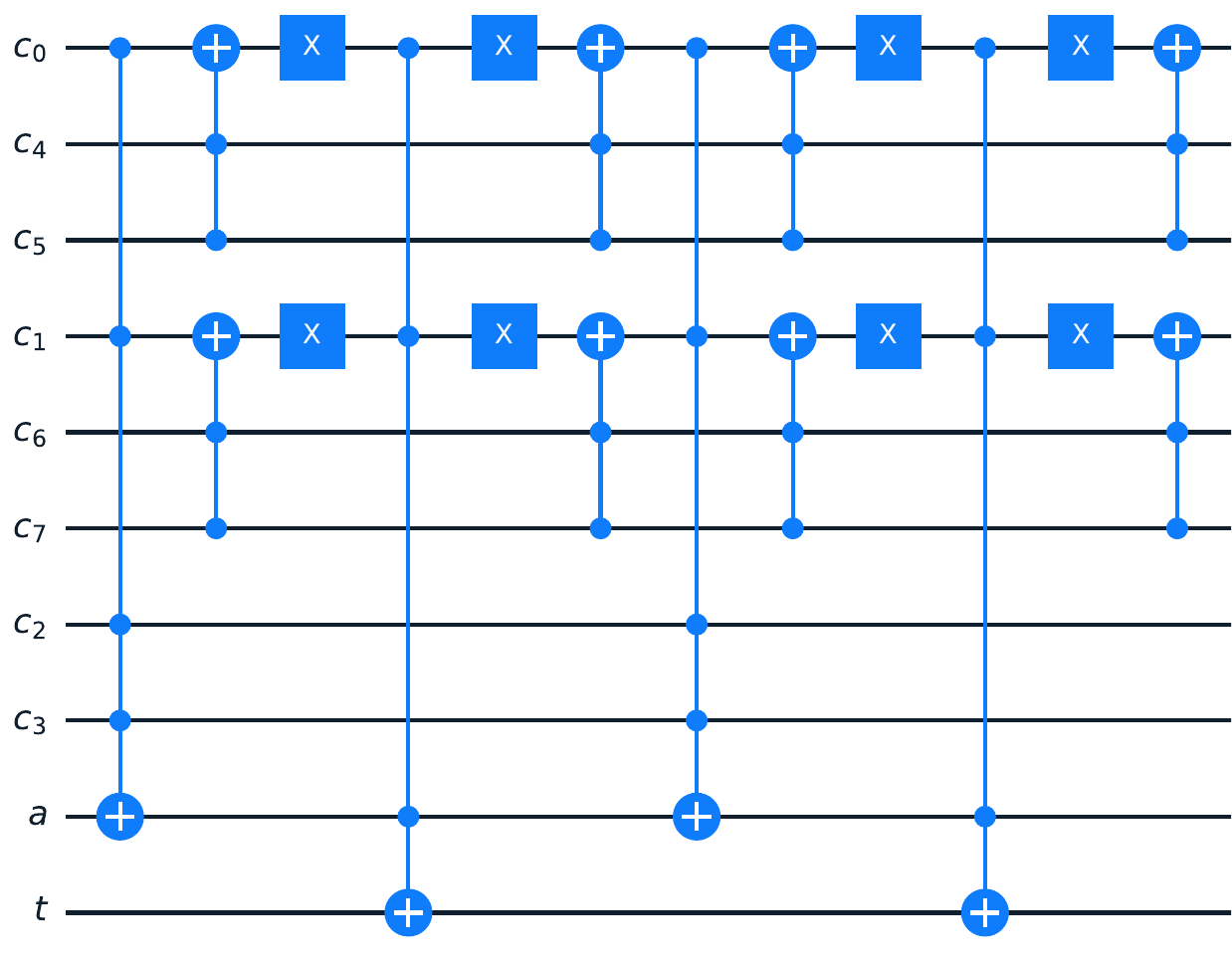}
        \caption{Log, Dirty.}
        \label{fig:single_aux:log}
    \end{subfigure}
    \caption{Single Aux decomposition for 8-controlled Pauli~\texttt{X}.}
    \label{fig:single_aux}
    \Description{}
\end{figure}

The \emph{Linear} algorithm relies on the application of a multi-con\-trolled $2\pi$ rotation gate into an auxiliary qubit, as illustrated in Figure~\ref{fig:single_aux:linear}. This results in a multi-controlled \texttt{Z} gate with target and control qubits, which can be transformed into a multi-controlled \texttt{X} via a basis change at the target. The decomposition of the multi-controlled rotation is performed without auxiliary with the SU(2) Linear algorithm~\cite{zindorfEfficientImplementationMultiControlled2024}.

The Single Aux Log decomposition~\cite{claudonPolylogarithmicdepthControlledNOTGates2024} recursively breaks an $n$-controlled \texttt{X} gate into  $2\sqrt{n}$-controlled \texttt{X} gates that can execute in parallel. The base case of the recursion is a 4-controlled \texttt{X} gate, which can be decomposed without auxiliary qubits. Unlike the Linear algorithm, the number of CNOTs and the circuit depth can be reduced when using a clean auxiliary qubit. Figure~\ref{fig:single_aux:log} illustrates the Single Aux Log decomposition of an 8-controlled \texttt{X} gate.

\subsubsection*{Linear Depth}

The Linear Depth decomposition algorithm proposed by \citet{dasilvaLineardepthQuantumCircuits2022}, as the name suggests, results in a circuit with linear depth. This is an efficient algorithm that requires no auxiliary qubits and can be used for any quantum gate. We consider this algorithm as the fallback decomposition since it has no requirements to be applied to any U(2) gate. However, this algorithm results in a circuit with a quadratic number of CNOTs.

\subsection{Rotation Gates}

Alongside the Network decomposition, Rotation gates allow for the use of the SU(2) decomposition algorithms, which requires no auxiliary qubits. We categorize these algorithms as SU(2) Linear, proposed by \citet{zindorfEfficientImplementationMultiControlled2024}, which is also used for the Single Aux Linear and SU(2) Rewrite decompositions~\cite{zindorfEfficientImplementationMultiControlled2024,rosaOptimizingGateDecomposition2025}; and SU(2) Log, which relies on the Single Aux Log decomposition algorithm for Pauli gates~\cite{claudonPolylogarithmicdepthControlledNOTGates2024}.

\begin{figure}[htbp]
    \centering
    \begin{subfigure}[b]{.53\linewidth}
        \centering
        \includegraphics[height=100px]{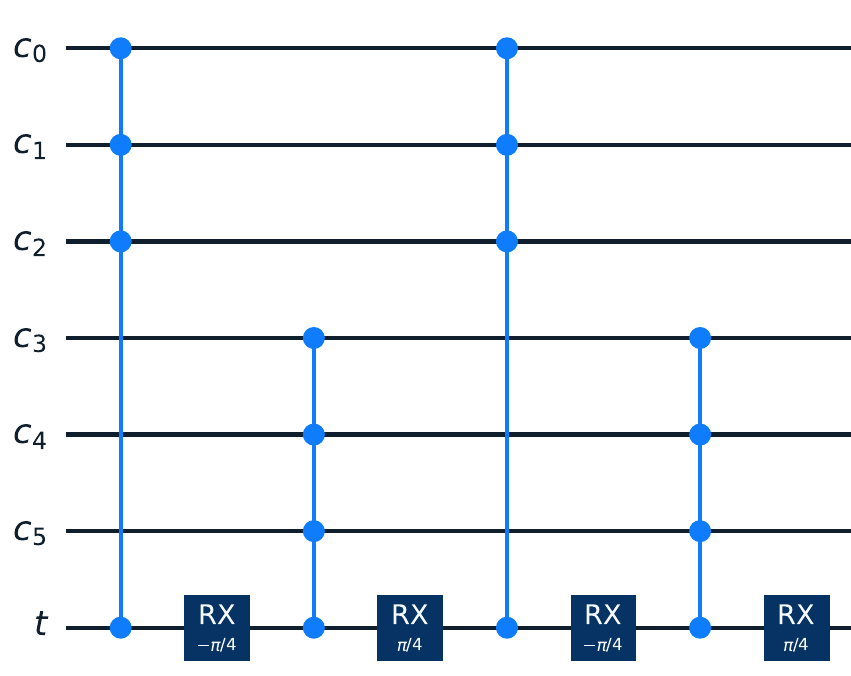}
        \caption{SU(2) Linear.}
        \label{fig:su2:linear}
    \end{subfigure}
    \hfil
    \begin{subfigure}[b]{.46\linewidth}
        \centering
        \includegraphics[height=100px]{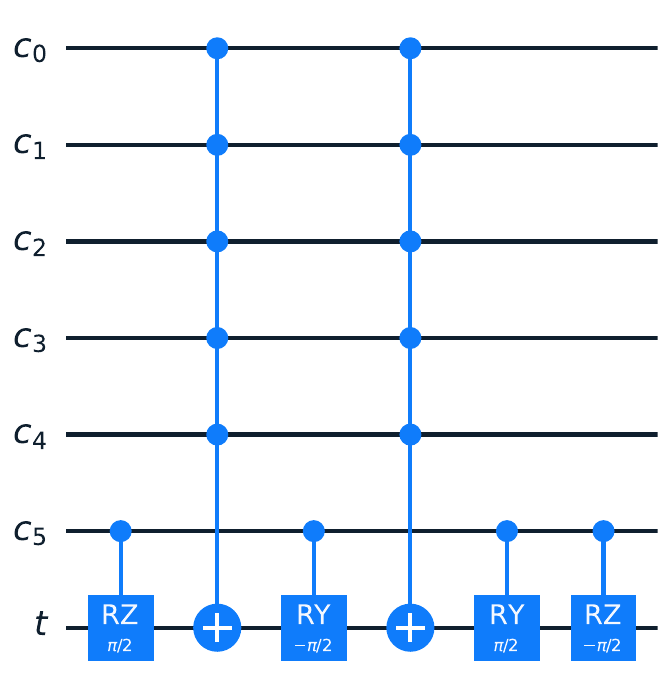}
        \caption{SU(2) Log.}
        \label{fig:su2:log}
    \end{subfigure}
    \caption{SU(2) decomposition for 6-controlled \texttt{RX($\pi$)}.}
    \label{fig:su2}
    \Description{}
\end{figure}

The SU(2) Linear decomposition algorithm uses an approximate version of a multi-controlled \texttt{Z} gate, as illustrated in Figure~\ref{fig:su2:linear}. Each multi-controlled \texttt{Z} gate is decomposed into an approximate version where free qubits serve as auxiliaries. The approximate decomposition of the \texttt{Z} gate relies on phase differences canceling out in pairs.

Figure~\ref{fig:su2:log} illustrates the decomposition using the SU(2) Log algorithm, where each multi-controlled \texttt{X} is decomposed using the Single Aux Log algorithm with a dirty auxiliary, and the single-controlled gates can be decomposed without an auxiliary qubit using just two CNOTs. Figure~\ref{fig:su2} illustrates the decomposition of a multi-controlled \texttt{RX}($\pi$) gate. For other Rotation gates and angles, the multi-controlled gates are arranged in the same structure.

\subsection{Phase and Hadamard Gates}

When multiple auxiliary qubits, or none, are available, the Network and Linear Depth decompositions can be used, respectively, as for most gates. However, when only a single auxiliary qubit is available, there are two decomposition algorithms available: SU(2) Rewrite~\cite{rosaOptimizingGateDecomposition2025}, which leverages the SU(2) Linear decomposition~\cite{zindorfEfficientImplementationMultiControlled2024}, and the Single Aux U(2), which uses the Single Aux Log decomposition~\cite{claudonPolylogarithmicdepthControlledNOTGates2024}.

\begin{figure}[htbp]
    \centering
    \begin{subfigure}[b]{.26\linewidth}
        \centering
        \includegraphics[height=112px]{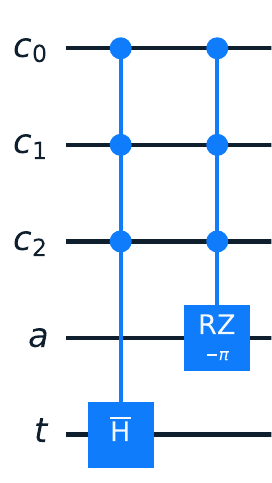}
        \caption{SU(2) Rewrite.}
        \label{fig:su2r:linear}
    \end{subfigure}
    \hfil
    \begin{subfigure}[b]{.26\linewidth}
        \centering
        \includegraphics[height=112px]{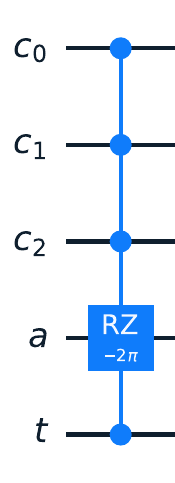}
        \caption{SU(2) Rewrite.}
        \label{fig:su2r:linear_p}
    \end{subfigure}
    \hfil
    \begin{subfigure}[b]{.37\linewidth}
        \centering
        \includegraphics[height=112px]{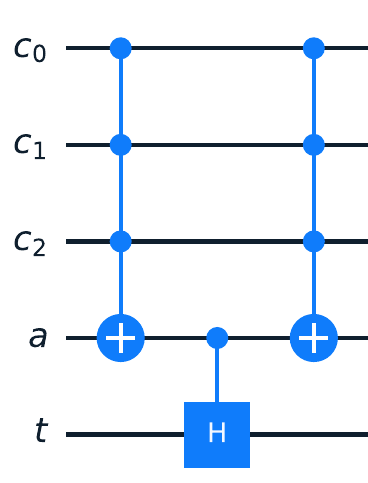}
        \caption{Single Aux U(2).}
        \label{fig:su2r:log}
    \end{subfigure}
    \caption{Decomposition for a 3-con\-trolled Ha\-da\-mard (\ref{sub@fig:su2r:linear}~and~\ref{sub@fig:su2r:log}) and a 3-con\-trolled Phase($\pi$) (\ref{sub@fig:su2r:linear_p}). In \ref{sub@fig:su2r:linear}, $\overline{\texttt{H}}$ is a SU(2) gate equivalent to \texttt{H}~\cite{rosaOptimizingGateDecomposition2025}.}
    \label{fig:su2r}
    \Description{}
\end{figure}

The SU(2) Rewrite decomposition relies on the fact that any U(2) gate can be represented by a SU(2) gate and a global phase, and that this global phase can be ``corrected'' by a multi-controlled \texttt{RZ} gate~\cite{rosaOptimizingGateDecomposition2025}. Figure~\ref{fig:su2r:linear} illustrates the decomposition of a multi-controlled Hadamard gate. For this decomposition, the two multi-controlled gates can be fused, resulting in the same complexity as decomposing a single multi-controlled SU(2) gate~\cite{zindorfEfficientImplementationMultiControlled2024}. For the multi-controlled Phase gate, illustrated in Figure~\ref{fig:su2r:linear_p}, the decomposition can be simplified, but this results only in a constant reduction in the number of CNOTs~\cite{zindorfEfficientImplementationMultiControlled2024}.

For the Single Aux U(2) decomposition, a clean auxiliary qubit is added to break the multi-controlled gate into two multi-controlled \texttt{X} gates and a dirty auxiliary, as illustrated in Figure~\ref{fig:su2r:log}. This decomposition results in a circuit of logarithmic depth since the multi-controlled \texttt{X} gates are then decomposed using the Single Aux Log decomposition. The decomposition method applied to the Phase and Hadamard gates can be extended to any other gate, but in many cases, a more efficient decomposition is available for Pauli and Rotation gates.

The decompositions listed for the U(2) gate in Table~\ref{tab:algorithms} are used for any gate as presented in the previous sections, with the exception of the Linear Depth decomposition for Rotation gates. In the next section, we present a benchmark of the decomposition algorithms discussed in this section.

\section{Algorithms Benchmark}
\label{sec:benchmark}

We implemented all the decomposition algorithms presented in Table~\ref{tab:algorithms} in the Ket quantum programming platform. This is the first implementation that consolidates the current state-of-the-art decomposition algorithms into a single platform, allowing us to collect numerical data from their execution. Our objective is to classify the performance of these algorithms in terms of compilation time and quantum execution time, creating compilation profiles where those key metrics will be taken into consideration. We classify the algorithms into two profiles: \emph{Compilation Time} and \emph{Execution Time}.

The \emph{Compilation Time} profile, as the name suggests, is optimized to reduce the classical execution time of the decomposition algorithms, which may also result in improved execution time for simulated quantum executions.
The \emph{Execution Time} profile targets large-scale Fault-Tolerant quantum computers~\cite{devittQuantumErrorCorrection2013,GoogleQuantumAIandCollaborators2025} with the potential to compute with millions of qubits.

As Ket's compiler automatically manages auxiliary qubits whenever they are available on the quantum computer~\cite{rosaAutomatedAuxiliaryQubit2024}, the data was gathered using high-level programming instructions such as ``\texttt{ctrl(c,~gate)(t)}'', where \texttt{c} is a list of control qubits, \texttt{gate} is a single-qubit gate, and \texttt{t} is the target qubit. These qubits belong to a quantum processor with enough additional qubits available for use as auxiliary in the decomposition. No actual quantum execution was performed during the tests; only the decomposition was evaluated.

The primary metrics we evaluated in our benchmarks are the compilation and quantum execution times required to decompose an $n$-controlled quantum gate. We measure the compilation time based on the number of CNOTs and the quantum execution time based on the quantum circuit depth. The decomposition of multi-qubit gates, which is the focus of this study, is just one step in compiling a quantum program. Therefore, measuring the time in seconds to compile the code or decompose the gates may not be a suitable metric for compilation time. Instead, the number of CNOTs allows us to infer the impact of the decomposition on subsequent compilation steps, which have time complexity directly related to the number of CNOTs.

We evaluate the quantum execution time in terms of quantum circuit depth. Since quantum gates that do not depend on each other can execute in parallel, the depth represents the minimum execution time required for the circuit. For the circuit depth metric, we consider only CNOT gates, as they have the most significant impact on quantum execution time. We disregard single-qubit gates, since sequences of such gates can be fused into a single operation, and in some cases, they can be virtually implemented~\cite{mckayEfficientGatesQuantum2017} with no impact on execution time. Also, the actual quantum execution time, in seconds, depends on hardware constraints such as qubit connectivity, which also affects the compilation step related to circuit mapping, as well as the availability of hardware resources to perform operations in parallel.

We organize our results using the same gate categories introduced in the previous section: Pauli gates, Rotation gates, and Phase and Hadamard gates. Next, we present the benchmark results for each group, followed by our analysis in Section~\ref{sec:results}.

\paragraph{Pauli Gates Benchmark}

\begin{figure}[htbp]
    \includegraphics[width=\linewidth]{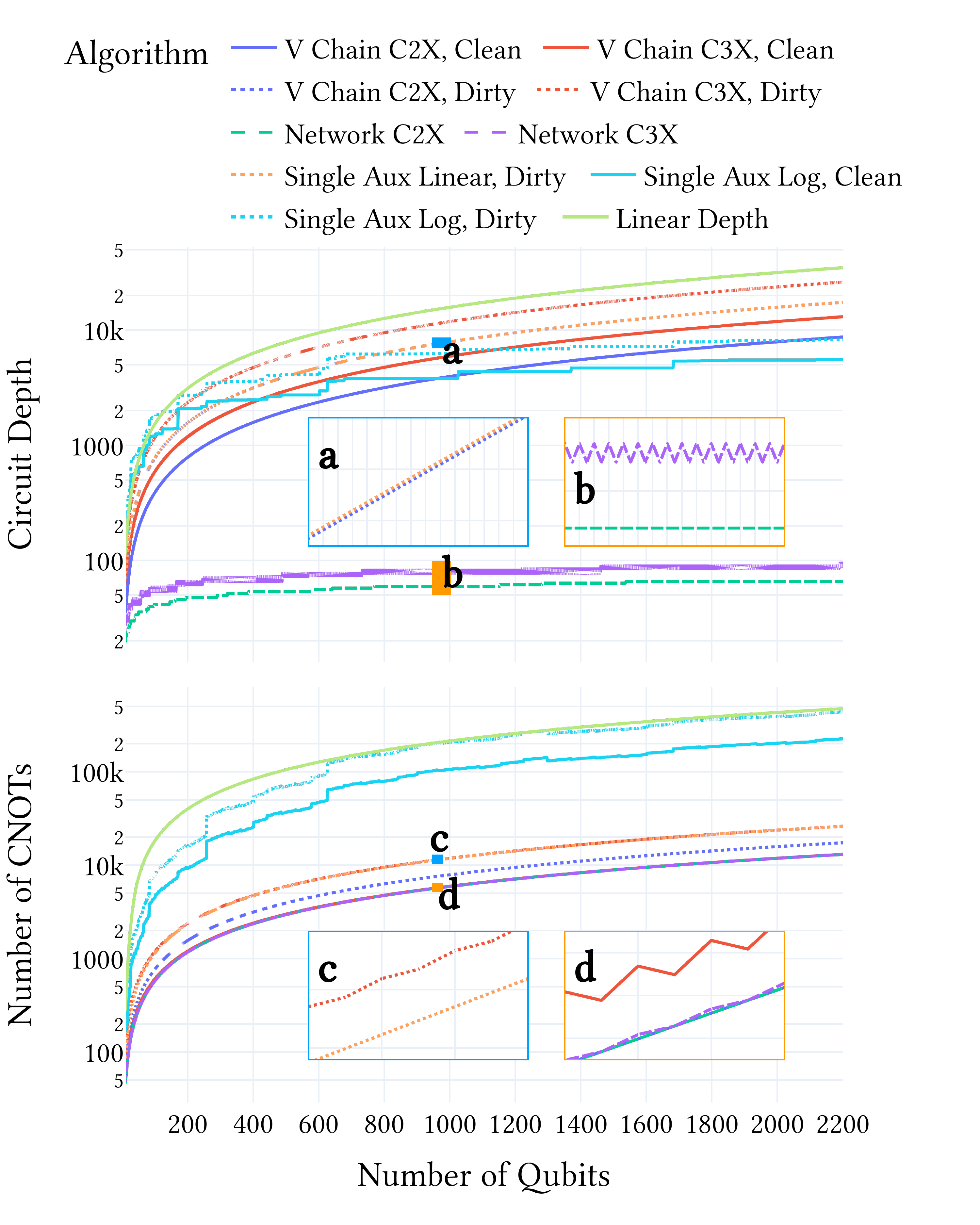}
    \caption{Comparison of various decomposition algorithms for \emph{Multi-Controlled Pauli Gates} across two metrics: Circuit Depth and Number of CNOTs. Insets display zoomed-in views for specific regions with matching colors.}
    \label{fig:result:pauli}
    \Description{}
\end{figure}

Figure~\ref{fig:result:pauli} presents the number of CNOTs and circuit depth for multi-controlled Pauli gates. The data was obtained with the instruction ``\texttt{ctrl(c,~X)(t)}'', but for the other Pauli gates, there is a difference of only a constant number of single-qubit gates, which does not affect the presented data.

\paragraph{Rotation Gates Benchmark}

\begin{figure}[htbp]
    \centering
    \includegraphics[width=\linewidth]{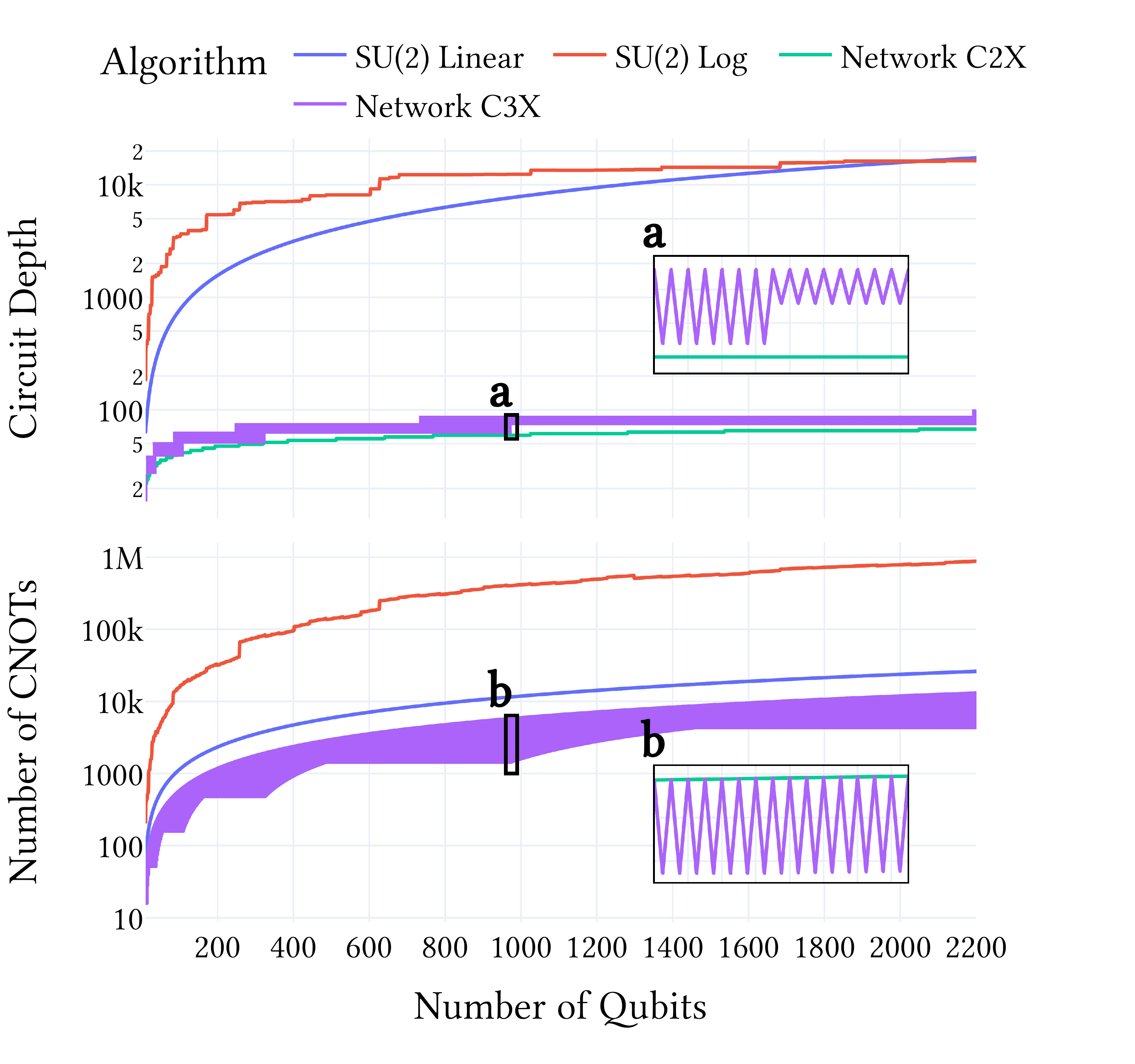}
    \caption{Comparison of various decomposition algorithms for \emph{Multi-controlled Rotation Gates} across two metrics: Circuit Depth and Number of CNOTs. Insets display zoomed-in views for specific regions.}
    \label{fig:result:rot}
    \Description{}
\end{figure}

Figure~\ref{fig:result:rot} presents the number of CNOTs and circuit depth for multi-controlled Rotation gates. The data was obtained using the instruction ``\texttt{ctrl(c,~RX(pi/2))(t)}'', but for the other Rotation gates, there is a difference of only a constant number of single-qubit gates, which does not affect the presented data.

\paragraph{Phase and Hadamard Gates Benchmark}

\begin{figure}[htbp]
    \centering
    \includegraphics[width=\linewidth]{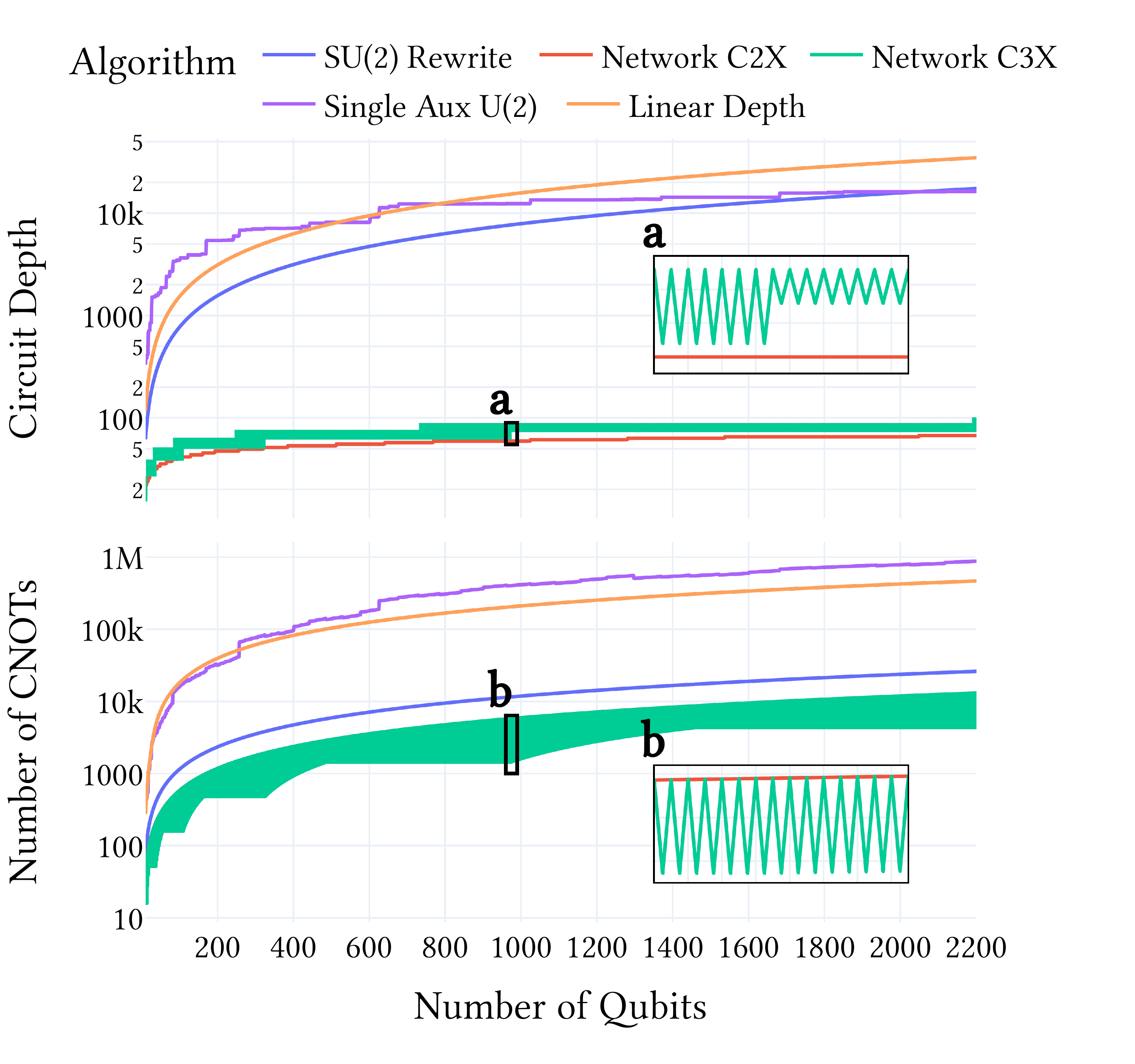}
    \caption{Comparison across different decomposition algorithms for \emph{Multi-controlled Phase and Hadamard Gates} across two metrics: Circuit Depth and Number of CNOTs. Insets display zoomed-in views for specific regions.}
    \label{fig:result:u2}
    \Description{}
\end{figure}

Figure~\ref{fig:result:u2} presents the number of CNOTs and circuit depth for multi-controlled Phase and Hadamard gates. The data was obtained with the instruction ``\texttt{ctrl(c,~H)(t)}'', but for the Phase gates, there is a difference of only a constant number of single-qubit gates, which does not significantly affect the data.

\section{Results Analysis}
\label{sec:results}

Based on the data presented in the previous section, Table~\ref{tab:profile} presents the quantum gate decomposition algorithms from Table~\ref{tab:algorithms}, classified into the two compilation profiles. Note that not all algorithms are included in Table~\ref{tab:profile}, as some decompositions with fewer auxiliary qubit requirements may be more efficient.

\begin{table}[htbp]
    \caption{Quantum gate decomposition algorithms ranked by priority for two compilation profiles.}
    \label{tab:profile}
    \footnotesize
    \begin{tabular}{cll}
        \toprule
        \multirow{2}{*}{Gate}           & \multicolumn{2}{c}{Compilation Profile}                                       \\
        \cmidrule(lr){2-3}
                                        & \multicolumn{1}{c}{Compilation Time}    & \multicolumn{1}{c}{Execution Time}  \\
        \midrule
        \midrule
        \multirow{6}{*}{Pauli Gates}    & $1^\text{st}$ Network C2X               & $1^\text{st}$ Network C2X           \\
                                        & $2^\text{nd}$ Network C3X               & $2^\text{nd}$ Network C3X           \\
                                        & $3^\text{rd}$ V Chain C3X, Clean        & $3^\text{rd}$ Single Aux Log, Clean \\
                                        & $4^\text{th}$ V Chain C2X, Dirty        & $4^\text{th}$ Single Aux Log, Dirty \\
                                        & $5^\text{th}$ Single Aux Linear         & $5^\text{th}$ Linear Depth          \\
                                        & $6^\text{th}$ Linear Depth              &                                     \\
        \cmidrule(lr){1-3}
        \multirow{3}{*}{Rotation Gates} & $1^\text{st}$ Network C3X               & $1^\text{st}$ Network C2X           \\
                                        & $2^\text{nd}$ SU(2) Linear              & $2^\text{nd}$ Network C3X           \\
                                        &                                         & $3^\text{rd}$ SU(2) Log             \\
        \cmidrule(lr){1-3}
        \multirow{4}{*}{\makecell{Phase and                                                                             \\Hadamard}} & $1^\text{st}$ Network C3X        & $1^\text{st}$ Network C2X                         \\
                                        & $2^\text{nd}$ SU(2) Rewrite             & $2^\text{nd}$ Network C3X           \\
                                        & $3^\text{rd}$ Linear Depth              & $3^\text{rd}$ Single Aux U(2)       \\
                                        &                                         & $4^\text{th}$ Linear Depth          \\
        \bottomrule
    \end{tabular}
\end{table}

The benchmark results reveal that the Network decomposition stands out as the most efficient algorithm across all gate types. However, this performance comes at the cost of requiring the largest number of clean auxiliary qubits. Interestingly, the initial assumption that increasing the number of auxiliary qubits would always lead to more efficient decompositions does not consistently hold. In some cases, additional auxiliaries provide no performance improvements.

Algorithms that produce logarithmic circuit depth demonstrate their advantages only in scenarios involving more than 1000 qubits. Despite their depth efficiency, these algorithms tend to perform poorly in terms of CNOT gate count, often ranking alongside the Linear Depth decomposition as the least efficient in this regard.

With the exception of Rotation gates--which can be decomposed without auxiliary qubits and still maintain a linear number of CNOTs--most other gate types fall back on the Linear Depth decomposition when no auxiliary qubits are available. This makes Linear Depth a crucial strategy for current quantum compilers. However, as future quantum devices provide access to more qubits, the relevance of this method may decline.

A particularly noteworthy result from the benchmarks is the significant performance gap between using clean versus dirty auxiliary qubits. The data suggest that having access to a quantum processor with twice the number of qubits as needed by the program can dramatically reduce both compilation time and quantum execution time. The Single Aux Linear algorithm is an exception, as its performance remains unaffected by the auxiliary qubit state.

\section{Final Remarks}
\label{sec:conclusion}

In this paper, we addressed the trade-off between compilation time and quantum execution time in quantum programs. We argue that this trade-off is more important for quantum applications than for classical ones, as the compilation of a quantum program cannot be performed \emph{a priori}.

Our analysis focuses on quantum gate decomposition, which is the first step in quantum code compilation. We have gathered data from the state-of-the-art quantum gate decomposition algorithms, which we implemented on the Ket quantum programming platform. To the best of our knowledge, this is the first implementation of all these algorithms within a single platform.

From the data presented in Section~\ref{sec:benchmark}, we created two compilation profiles: one focused on reducing compilation time and the other on reducing quantum execution time. These results are presented in Table~\ref{tab:profile}. The quantum execution time profile considers large-scale quantum computers that go beyond current capabilities but will be necessary to run algorithms such as Shor's algorithm~\cite{shorPolynomialTimeAlgorithmsPrime1997}, which could break RSA encryption~\cite{gidneyHowFactor20482021}. Additionally, the presented data can be used to construct a lookup table that selects the best decomposition algorithm based on the number of qubits involved in a multi-qubit gate, as the performance may not be easily predictable for operations with fewer than 2200 qubits.

The results of this paper can help quantum compiler developers select the best decomposition algorithms for their compilers, depending on the type of gate and the number of control qubits. Note that the selection of the decomposition algorithm can be automatically performed by the compiler based on the available auxiliary qubits~\cite{rosaAutomatedAuxiliaryQubit2024}. However, the findings from this paper can also aid quantum developers in identifying the performance associated with each multi-controlled gate, enabling them to make more informed decisions when selecting instructions to use.

In our study, we only evaluated the decomposition step of quantum compilation. However, circuit mapping may have a significant impact on the final performance of the quantum program. Circuit mapping adjusts the quantum program to the connectivity constraints of the quantum computer. This means that, regardless of the circuit mapping algorithm used, different quantum computer architectures (or qubit connectivity structures) may be better suited to different decomposition algorithms. Future work could analyze these decomposition algorithms across various qubit connectivity structures, such as line, grid, and torus configurations.

Another direction for future work is to analyze the performance of the decomposition algorithms in the context of non-Clifford gates~\cite{aaronsonImprovedSimulationStabilizer2004}, as these gates have a significant impact on quantum execution time when the program is encoded in a Quantum Error Correction code~\cite{piveteauErrorMitigationUniversal2021}. Additionally, examining the impact of circuit optimization techniques, such as ZX-calculus~\cite{Coecke2011}, on the final circuit would be an interesting direction for further research.

\section*{ARTIFACT AVAILABILITY}

The authors declare that the research artifacts supporting the findings of this study
are accessible at \url{https://doi.org/10.5281/zenodo.16964846}.

\section*{ACKNOWLEDGMENTS}

ECRR acknowledges the Coordenação de Aperfeiçoamento de Pessoal de Nível Superior - CAPES, Finance Code 001; EID, JM, and ECRR acknowledges the Conselho Nacional de De\-sen\-vol\-vi\-men\-to Ci\-en\-tí\-fi\-co e Tec\-no\-ló\-gi\-co - CNPq through grant number \mbox{409673/2022-6}; JM, EID, and ECRR acknowledges the Fundação de Amparo à Pesquisa e I\-no\-va\-ção do Estado de Santa Catarina - FAPESC through Project FAPESC TR nº 2024TR002672.

\bibliographystyle{ACM-Reference-Format}
\bibliography{main}

\end{document}